# Toward The Universal Quantum Cloner Limit For Designing Compact Photonic CNOT Gate

Amor Gueddana[1,2*] and Vasudevan Lakshminarayanan [2,3]

*1 Green & Smart Communication Systems Lab, Gres'Com, Engineering School of Communication of Tunis, Sup'Com, University of Carthage, Ghazela Technopark, 2083, Ariana, Tunisia.*
*2 Theoretical & Experimental Epistemology Lab, TEEL, School of Optometry and Vision Science, University of Waterloo 200 University Avenue West, Waterloo, Ontario N2l 3G1, Canada.*
*3 Department of Physics, Department of Electrical and Computer Engineering and Department of Systems Design Engineering , University of Waterloo 200 University Avenue West, Waterloo, Ontario N2l 3G1, Canada.*

*E-mail: amor.gueddana@supcom.tn

**Abstract:** We suggest a near deterministic compact model of a photonic CNOT gate based on a quantum dot trapped in a double sided optical microcavity and a universal cloner. Our design surpasses the cloner optimal limit of 5/6 and we show that it provides fidelity around 91 % in the weak coupling regime.

## 1. Introcuction

Small scale and fully programmable Quantum computers based on trapped ions and superconducting circuits have emerged recently [1]. Nonetheless, quantum computers based on photonic qubits have not yet been realized. Working with photonic qubits has the advantage of being in lowest interaction with the environment in order to save superposition and entanglement of states. Unfortunately to date, photonic gates constituting the photonic quantum circuits are still non-deterministic. The universal quantum gate for building such circuits is known as the Controlled-NOT (CNOT) gate. The best photonic CNOT encountered in the literature is discussed in [2], and provides a theoretical fidelity of 93.7 %. This CNOT is based on spin of a single electron in a Quantum Dot (QD) trapped in a double sided optical microcavity which behaves like a Beam Splitter (BS). For this model, we presented demonstrations that it performs correct CNOT function only in the strong coupling regime [3]. In another work [4], we proposed an optimized version based on Universal Cloner (UC) and providing a fidelity of 78 % with better configuration of the outputs and independent of spin measure. In this paper, we propose a compact version of this CNOT and show that it can exceed the UC optimal cloning limit of 5/6 =83% [5, 6].
This paper has four sections: section 2 details modeling method of our compact CNOT and gives general description of realizing the CNOT function. In section 3, we present our simulation results. Section 4 is dedicated to the conclusion and future perspectives.

## 2. Compact CNOT Model

The compact CNOT model proposed in this work is illustrated in Fig. 1. It is composed of three Circular Polarizing Beam Splitters (CPBS), denoted $CPBS_1$, $CPBS_2$ and $CPBS_3$, a UC with fidelity denoted $F_{UC} < 1$, two single qubit gates ( $P_\phi$ and $\sigma_x$ detailed later), two Delay Lines (DL) , denoted $DL_1$ and $DL_2$, a BS 50:50, a photonic circulator and a QD system.
Logical 0 and 1 values are physically represented by right-circularly polarized and left-circularly polarized single photons, denoted $|R\rangle$ and $|L\rangle$, respectively. Our CNOT is based on GaAs/InAs QD system with single electron trapped in a double sided optical microcavity and being initially at the spin up state:

$$|\psi_s\rangle = |\uparrow_s\rangle \tag{1}$$

All the dynamics of the QD is described in [2, 4, 7], the transmission and reflection coefficients of the double sided optical micro-cavity system, denoted by $t(\omega)$ and $r(\omega)$, in the case of the coupled cavity, and for equal frequencies of the input photon, cavity mode and the spin-dependent optical transition, are:

$$\mathrm{t}(\omega)=\frac{2\rho\kappa}{\rho(2\kappa+\kappa_s)+4g^2};\ \mathrm{r}(\omega)=1+\mathrm{t}(\omega) \tag{2}$$

where $\kappa$ and $\kappa_s/2$ are the cavity field decay rate into the input/output modes and the leaky modes, respectively, $g$ is the coupling strength, and $\rho/2$ represents the $X^-$ dipole decay rate. In the case where the cavity is uncoupled, the transmission and reflection coefficients are denoted by $\mathrm{t}_0(\omega)$ and $\mathrm{r}_0(\omega)$, and can be derived from equation 2 with $g=0$. The interaction of the QD spin, when considering bit-flip errors that might be introduced, for $\mathrm{t}_0=|\mathrm{t}_0(\omega)|$, $\mathrm{r}_0=|\mathrm{r}_0(\omega)|$, $\mathrm{t}_1=|\mathrm{t}(\omega)|$ and $\mathrm{r}_1=|r(\omega)|$, are given as follows [2, 4]:

$$\begin{aligned}&|R^{\downarrow},\uparrow_s\rangle\rightarrow -t_0|R^{\downarrow},\uparrow_s\rangle-r_0|L^{\uparrow},\uparrow_s\rangle;\ |R^{\downarrow},\downarrow_s\rangle\rightarrow r_1|L^{\uparrow},\downarrow_s\rangle+t_1|R^{\downarrow},\downarrow_s\rangle\\&|L^{\uparrow},\uparrow_s\rangle\rightarrow -t_0|L^{\uparrow},\uparrow_s\rangle-r_0|R^{\downarrow},\uparrow_s\rangle;\ |L^{\uparrow},\downarrow_s\rangle\rightarrow r_1|R^{\downarrow},\downarrow_s\rangle+t_1|L^{\uparrow},\downarrow_s\rangle\end{aligned} \tag{3}$$

where $R^{\uparrow}$ and $R^{\downarrow}$, stand for a single photon entering the QD from the lower side (see b in Fig. 1) and upper side (see a in Fig. 1), respectively (same notations for $L^{\uparrow}$ and $L^{\downarrow}$).
The compact CNOT has two input single photons for the control and target qubits, it uses the UC to copy the polarization degree of freedom of the control qubits in order to control the target qubit, only when the latter is in the state $|L\rangle$.

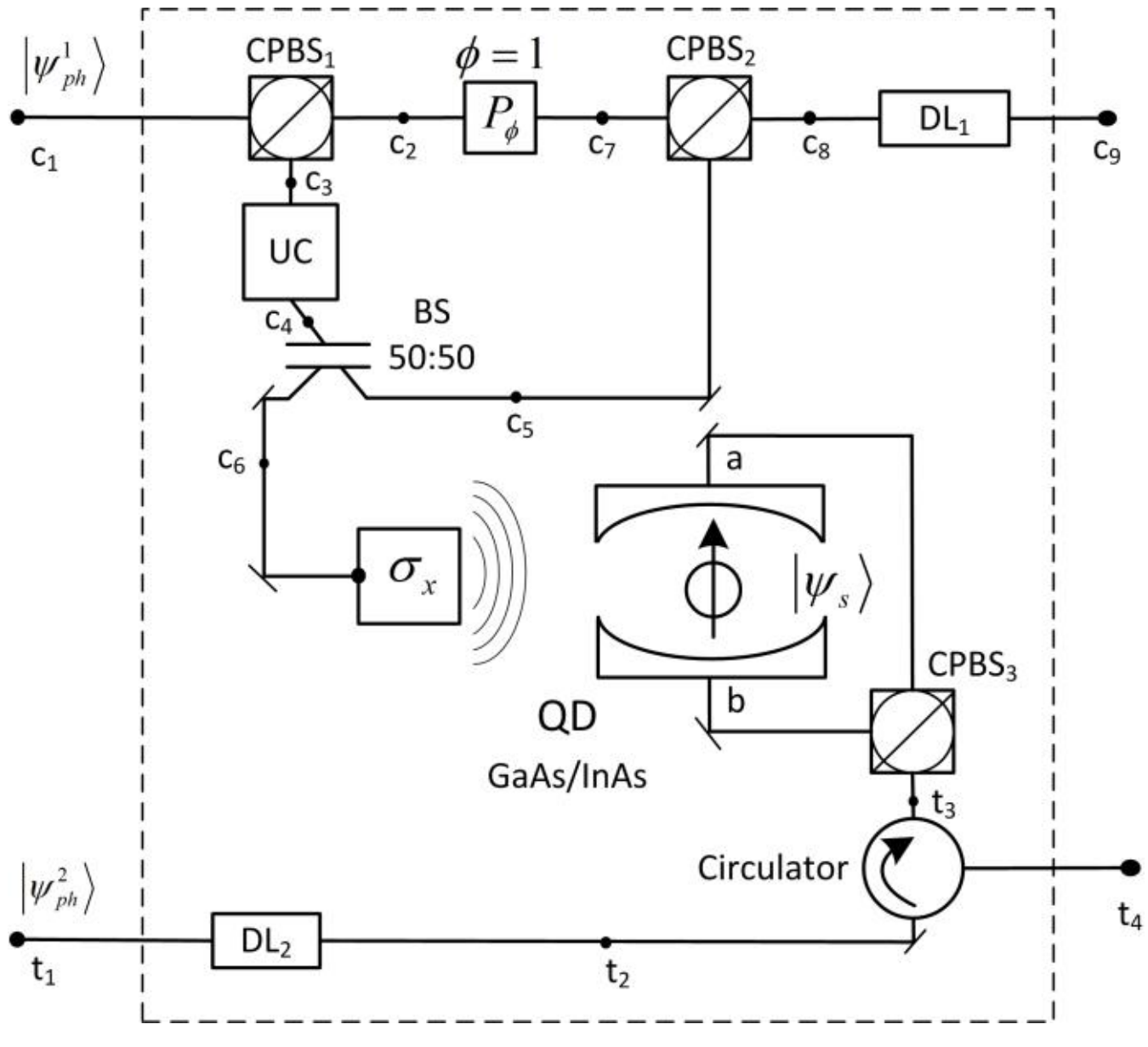

**Figure 1** Compact CNOT Gate using the universal cloner.

Let us consider photon 1 and photon 2, representing control and target, and having quantum states denoted $|\psi^1_{ph}\rangle_{c_1}$ and $|\psi^2_{ph}\rangle_{t_1}$, where $c_1$ and $t_1$ refer to control and target qubit, being at the specific points $c_1$ and $t_1$ marked in Fig. 1. To describe the action of each component of the compact CNOT, we use the same notations for all $\{c_i\}_{1\le i\le 9}$ and $\{t_j\}_{1\le j\le 4}$.

$|\psi^1_{ph}\rangle_{c_1}$ and $|\psi^2_{ph}\rangle_{t_1}$ are initially written as :

$$|\psi^1_{ph}\rangle_{c_1}=\alpha|R_1\rangle+\beta|L_1\rangle;\ |\psi^2_{ph}\rangle_{t_1}=\delta|R_2\rangle+\gamma|L_2\rangle \tag{3}$$

After travelling CPBS$_1$, we have:

$$|\psi^1_{ph}\rangle_{c_1}\xrightarrow{CPBS_1}|\psi^1_{ph}\rangle_{c_2}=\alpha|R_1\rangle\quad;\quad|\psi^1_{ph}\rangle_{c_3}=\beta|L_1\rangle \tag{4}$$

The single qubit gate $P_\phi$ in the circuit, where $\phi \in \mathbb{R}$, has the following general matrix expression:

$$U_{P_\phi} = \begin{pmatrix} e^{i\phi\pi} & 0 \\ 0 & e^{i\phi\pi} \end{pmatrix} \tag{5}$$

The value $\phi = 1$ is used in our compact CNOT, in order to introduce a negative sign shift to the state $\left|\psi_{ph}^1\right\rangle_{c_2}$ as follows:

$$\left|\psi_{ph}^1\right\rangle_{c_2} \xrightarrow{P_1} \left|\psi_{ph}^1\right\rangle_{c_7} = -\alpha\left|R_1\right\rangle \tag{6}$$

For the control photon being in the state $\left|\psi_{ph}^1\right\rangle_{c_3} = \beta\left|L_1\right\rangle$, and after passing through the UC having a fidelity $F_{UC}$, it becomes:

$$\left|\psi_{ph}^1\right\rangle_{c_3} \xrightarrow{UC} \left|\psi_{ph}^1\right\rangle_{c_4} = 2\sqrt{F_{UC}}\left|\psi_{ph}^1\right\rangle_{c_3} = 2\sqrt{F_{UC}}\beta\left|L_1\right\rangle \tag{8}$$

The BS 50:50 is used to separate the beam from UC and we have:

$$\left|\psi_{ph}^1\right\rangle_{c_4} \xrightarrow{BS\ 50:50} \left|\psi_{ph}^1\right\rangle_{c_5} = \left|\psi_{ph}^1\right\rangle_{c_6} = \sqrt{F_{UC}}\beta\left|L_1\right\rangle \tag{9}$$

CPBS$_2$ transmits $\left|\psi_{ph}^1\right\rangle_{c_7}$ and reflects $\left|\psi_{ph}^1\right\rangle_{c_5}$:

$$\left|\psi_{ph}^1\right\rangle_{c_7} + \left|\psi_{ph}^1\right\rangle_{c_5} \xrightarrow{CPBS_2} \left|\psi_{ph}^1\right\rangle_{c_8} = -\alpha\left|R_1\right\rangle + \sqrt{F_{UC}}\beta\left|L_1\right\rangle \tag{7}$$

DL$_1$ is then used to wait for the interaction of the cloned control photon with the QD system and target qubits. The single photon present at the state $\left|\psi_{ph}^1\right\rangle_{c_6}$ is used to control the single qubit $\sigma_x$ transform performed on the spin electron state, through $\pi$ microwave pulses [8, 9], which transforms the spin states as follows:

$$\left|\psi_s\right\rangle = \left|\uparrow_s\right\rangle \rightarrow \left|\downarrow_s\right\rangle \ ; \ \left|\psi_s\right\rangle = \left|\downarrow_s\right\rangle \rightarrow \left|\uparrow_s\right\rangle \tag{10}$$

This $\sigma_x$ transform is required only when the control qubit is in the state $\left|L_1\right\rangle$, and it is performed twice, exactly before and after the target qubits travels the QD system.
Concerning the target, and after having been delayed by DL2 to wait for control photon copying, we have $\left|\psi_{ph}^2\right\rangle_{t_1} = \left|\psi_{ph}^2\right\rangle_{t_2}$. The photonic circulator used in Fig.1 is necessary to separate the input photon (travelling from $t_2$ to $t_3$) from the output photon (coming from $t_3$ to the output $t_4$), therefore we have $\left|\psi_{ph}^2\right\rangle_{t_3} = \left|\psi_{ph}^2\right\rangle_{t_2}$ before entering the QD system. After interaction of $\left|\psi_{ph}^2\right\rangle_{t_3}$ with CPBS$_3$ and the QD system, our compact CNOT transforms the intial states given by equations 1 and 3 to the following:

$$\begin{gathered}
\left|\psi_{ph}^1\right\rangle_{c_1} \otimes \left|\psi_{ph}^2\right\rangle_{t_1} \otimes \left|\psi_s\right\rangle \rightarrow \left|\psi_{ph}^1\right\rangle_{c_9} \otimes \left|\psi_{ph}^2\right\rangle_{t_4} \otimes \left|\psi_s\right\rangle \\
= \left(\eta_1\left|R_1, R_2\right\rangle + \eta_2\left|R_1, L_2\right\rangle + \eta_3\left|L_1, L_2\right\rangle + \eta_4\left|L_1, R_2\right\rangle\right) \otimes \left|\uparrow_s\right\rangle \\
\text{where} \\
\eta_1 = \alpha\left(t_0\delta + r_0\gamma\right);\ \eta_2 = \alpha\left(r_0\delta + t_0\gamma\right) \\
\eta_3 = \sqrt{F_{UC}}\beta\left(r_1\delta + t_1\gamma\right);\ \eta_4 = \sqrt{F_{UC}}\beta\left(t_1\delta + r_1\gamma\right)
\end{gathered} \tag{11}$$

## 3. Simulation Results

To compare the fidelity of this compact CNOT to the ideal, we use the fidelity as proposed in [2]. The strong coupling regime is obtained for $g > \left(\kappa_s + \kappa\right)/4$ and the weak coupling regime is obtained for $g < \left(\kappa_s + \kappa\right)/4$. We simulated the average fidelity of our CNOT versus the normalized coupling strengths $\kappa_s/\kappa$ and $g/\kappa$ as illustrated by Fig. 2. We consider for our simulation the theoretical optimal fidelity value of the UC given by $F_{UC} = 5/6$.

Highest fidelity values for our compact CNOT are obtained only in the weak coupling regime, and for the specific values $\kappa_s/\kappa = g/\kappa = 0.01$, we reach a maximum of CNOT fidelity equal to $\overline{F_{CNOT}} = 90,43\%$.

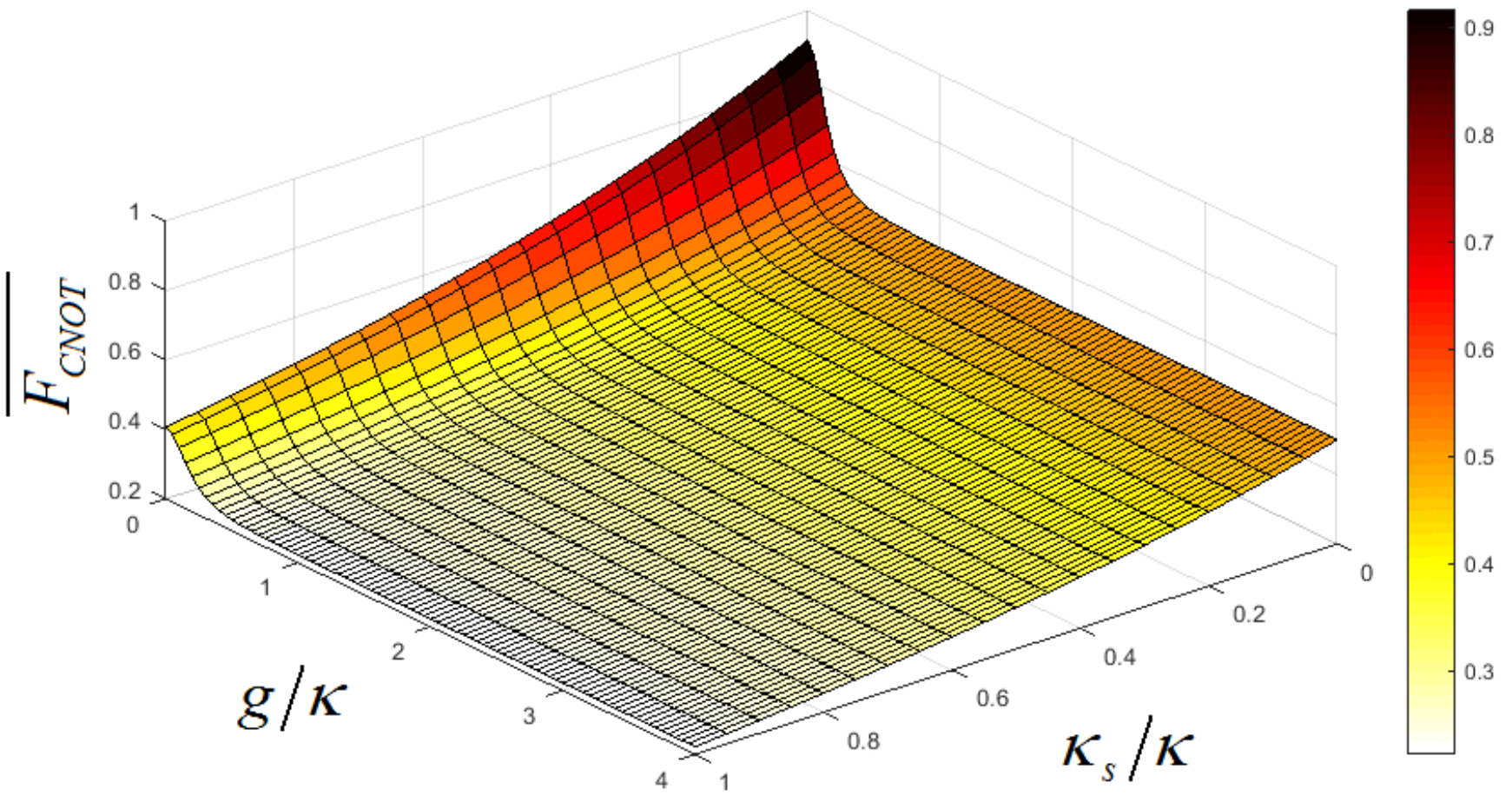

**Figure 2.** The average fidelity of the photonic CNOT gate versus the normalized coupling strengths $\kappa_s/\kappa$ and $g/\kappa$. $\rho$ is set to the specific value $\rho = 0.1\kappa$.

## 4. Conclusion

We demonstrated a compact CNOT model capable of providing correct CNOT functioning with fidelity around 91 %. Even if we have slightly surpassed the theoretical optimal limit of the quantum cloner for designing a CNOT with photonic qubits, still much work is necessary in order to obtain deterministic CNOTs which are physically realizable with low error probability. This implementation challenges will allow universal photonic quantum computer to be available in order to evaluate complex quantum algorithms such as database search and QFT.